\documentclass[onecolumn]{autart}    

\usepackage{graphicx}          

\usepackage{graphics} 
\usepackage{epsfig} 
\usepackage{mathptmx} 
\usepackage{times} 
\usepackage{amsmath} 
\usepackage{amssymb}  
\usepackage{floatrow}
\usepackage{multirow}

\begin{document}

\begin{frontmatter}

\title{Quantum gate identification: error analysis, numerical results and optical experiment\thanksref{footnoteinfo}} 

\thanks[footnoteinfo]{This work was supported by the Australian Research Council's Discovery Projects funding scheme under Project DP130101658, Laureate Fellowship FL110100020, AFOSR under grant FA2386-16-1-4065, Centres of Excellence CE110001027, NSFC 61374092, 61621003 and National Key Basic Research Program of China (973 program) under grant No. 2014CB845302.}
\author[adfa,cqcct]{Yuanlong Wang}\ead{yuanlong.wang.qc@gmail.com},    
\author[ustc1,ustc2]{Qi Yin}\ead{yinqi@mail.ustc.edu.cn},               
\author[adfa,b]{Daoyi Dong}\ead{daoyidong@gmail.com},  
\author[cas,ucas]{Bo Qi}\ead{qibo@amss.ac.cn},
\author[anu]{Ian R. Petersen}\ead{i.r.petersen@gmail.com},
\author[ustc1,ustc2]{Zhibo Hou}\ead{houzhibo@ustc.edu.cn},
\author[adfa,cqcct]{Hidehiro Yonezawa}\ead{h.yonezawa@adfa.edu.au},
\author[ustc1,ustc2]{Guo-Yong Xiang}\ead{gyxiang@ustc.edu.cn}
\thanks[b]{Tel. +61-2-62686285, Fax +61-2-62688443 (Daoyi Dong).}

\address[adfa]{School of Engineering and Information Technology, University of New South Wales, Canberra, ACT 2600, Australia}  
\address[cqcct]{Centre for Quantum Computation and Communication Technology, Australian Research Council, Canberra, ACT 2600, Australia}             
\address[ustc1]{Key Laboratory of Quantum Information, University of Science and Technology of China, CAS, Hefei 230026, People's Republic of China}        
\address[ustc2]{Synergetic Innovation Center of Quantum Information and Quantum Physics, University of Science and Technology of China, Hefei, Anhui 230026, People's Republic of China}
\address[cas]{Key Laboratory of Systems and Control, Academy of Mathematics and Systems Science, CAS, Beijing 100190, People's Republic of China}
\address[ucas]{University of Chinese Academy of Sciences, Beijing 100049, People's Republic of China}
\address[anu]{Research School of Engineering, Australian National University, Canberra, ACT 0200, Australia}


\begin{abstract}                          
The identification of an unknown quantum gate is a significant issue in quantum technology. In this paper, we propose a quantum gate identification method within the framework of quantum process tomography. In this method, a series of pure states are inputted to the gate and then a fast state tomography on the output states is performed and the data are used to reconstruct the quantum gate. Our algorithm has computational complexity $O(d^3)$ with the system dimension $d$. The algorithm is compared with maximum likelihood estimation method for the running time, which shows the efficiency advantage of our method. An error upper bound is established for the identification algorithm and the robustness of the algorithm against the purity of input states is also tested. We perform quantum optical experiment on single-qubit Hadamard gate to verify the effectiveness of the identification algorithm.
\end{abstract}

\end{frontmatter}

\section{Introduction}\label{Secintro}
In the recent decades considerable efforts have been devoted to the research of quantum technology. The identification of quantum systems is pivotal to tasks like establishing quantum communication networks and building practical quantum computers \cite{burgarth 2012}. Topics in quantum system identification include identifiability of quantum systems, single parameter identification, quantum Hamiltonian identification (\cite{zhang 2014}, \cite{leghtas 2012}, \cite{my 2017}, \cite{bris 2007}, \cite{bonnabel 2009}, \cite{sone 2017}), structure identification, mechanism identification \cite{Shu 2017}, quantum process identification, etc. For example, Burgarth and Yuasa \cite{burgarth 2012} established a general framework to classify the amount of attainable knowledge of a quantum system from a given experimental setup. Gu\c{t}\u{a} and Yamamoto \cite{guta 2016} presented a criterion for identifiability of passive linear quantum systems and two concrete identification methods. They also used Fisher information to optimize input states and output measurements. Kato and Yamamoto \cite{kato 2014} designed a continuous-time Bayesian update method to identify an unknown spin network structure with a high probability. They also employed an adaptive measurement technique to deterministically drive any mixed state to a spin coherent initial state. Sone and Cappellaro \cite{sone 2017} employed Gr\"{o}bner basis to determine the identifiability of many-body spin-half systems assisted by a single quantum probe. A series of quantum process tomography (QPT) methods (\cite{chuang 1997}, \cite{poyatos 1997}, \cite{jezek 2003}, etc.) have been developed to fully identify an unknown quantum process.

In this paper, we focus on the identification of quantum gates. As the quantum edition analogy of classical logical gates, quantum gates are fundamental tools to manipulate qubits and to implement quantum logical operations. Therefore, they are essential components for quantum information and quantum computation, and the task of identifying an unknown quantum gate is vital to verify and benchmark quantum circuits and quantum chips \cite{Nielsen and Chuang 2000}.

A natural approach to quantum gate identification is to view the unitary gate as a special class of quantum process. Many results have been obtained from this point of view. Gutoski and Johnston \cite{gutoski 2014} proved that any $d$-dimension unitary channel can be determined with only $O(d^2)$ interactive observables. Baldwin \emph{et al.} \cite{baldwin 2014} showed that a $d$-dimension unitary map is completely characterized by a minimal set of $d^2+d$ measurement outcomes and need to be achieved using at least $d$ probe pure states. Wang \emph{et al.} \cite{wang 2016} proposed an adaptive unitary process tomography protocol which needs only $d^2+d-1$ measurement outcomes for a $d$-dimension system. For the quantum gates having an efficient matrix product operator representation, Holz\"{a}pfel \emph{et al.} \cite{youyong2015} presented a tomography method that requires only measurements of linearly many local observables on the subsystems. Furthermore, standard quantum process tomography methods can be used to identify an unknown quantum gate. For example, Maximum Likelihood Estimation (MLE) for QPT has been applied to gate identification in \cite{brien 2004}, \cite{micuda 2015}, \cite{beterov 2016}. Bayesian deduction method for QPT has also been applied to gate identification in \cite{teklu}.

There are also other methods to solve the gate identification problem. Kimmel \emph{et al.} \cite{kimmel} developed a parameter estimation technique to calibrate key systematic parameters in a universal single-qubit gate set and achieved good robustness and efficiency. Rodionov \emph{et al.} \cite{rodionov 2014} utilized compressed sensing QPT to characterize quantum gates based on superconducting Xmon and phase qubits. They showed that it is of high probability that compressed sensing method can reduce the amount of resources significantly. Kimmel \emph{et al.} \cite{kimmel 2016} used randomized benchmarking method to reconstruct unitary evolution and achieved robustness to preparation, measurement and gate imperfections. For these existing methods, it is usually difficult either to characterize the computational complexity or to perform error analysis theoretically. This paper proposes a novel identification algorithm for quantum gates, analyzes the computational complexity and establishes an error upper bound.

For a unitary quantum gate, the output state is always a pure state if the input state is a pure state. In this paper, we take advantage of this property to present a novel algorithm for gate identification. The identification method is introduced within the QPT framework in \cite{Nielsen and Chuang 2000} and based on the result on quantum Hamiltonian identification in \cite{my 2016}. In \cite{my 2016}, a quantum Hamiltonian identification method has been presented and the algorithm has computational complexity $O(d^6)$ with the system dimension $d$. For our method in this paper, first a series of determined probe quantum pure states are inputted to the quantum gate, then the output states are measured and the gate is identified using the measurement data. It is proved that our method has computational complexity $O(d^3)$, which is much lower than the complexity $O(d^6)$ in \cite{my 2016}. We also demonstrate the expectation of error has a scaling of $O(\frac{d^{2.5}}{\sqrt{N}})$ where $N$ is the total resource number of input states. Our numerical results are in accordance with the theoretical error analysis. Furthermore, we perform quantum optical experiment on one-qubit Hadamard gate to demonstrate the theoretical result.

The structure of this paper is as follows. In Section \ref{Sec2} we rephrase the QPT framework in \cite{Nielsen and Chuang 2000} and formulate the gate identification problem. Section \ref{Sec3} presents the quantum gate identification algorithm. Section \ref{secadd} analyzes the computational complexity of our method and proves an error upper bound. Section \ref{Sec4} presents numerical results to demonstrate the performance of the proposed method. Section \ref{SecExp} illustrates quantum optical experimental results. Section \ref{secfinal} concludes this paper.

\section{Preliminaries and problem formulation}\label{Sec2}
\subsection{Quantum system}
The state of a finite-dimensional closed quantum system can be represented by a unit complex vector $|\psi\rangle$ and its dynamic evolution can be described by the Schr\"{o}dinger equation
\begin{equation}
\text{i}\frac{\partial}{\partial t}|\psi(t)\rangle=H|\psi(t)\rangle,
\end{equation}
where $H$ is the system Hamiltonian, $\text{i}=\sqrt{-1}$ and we use atomic units to set $\hbar=1$ in this paper. $|\psi\rangle$ is also called a pure state. The probabilistic mixture of pure states is called a mixed state, which is described by a density matrix $\rho$. For pure states, $\rho=|\psi\rangle\langle\psi|$ and $\text{Tr}(\rho^2)=1$, while in the general case $\rho$ is a Hermitian, positive semidefinite matrix satisfying $\text{Tr}(\rho)=1$ and $\text{Tr}(\rho^2)\leq1$. To make measurements on a state, a set $\{P_i\}$ of positive operator valued measurement (POVM) elements are prepared, where the elements $P_i$ are positive semidefinite Hermitian matrices and sum to identity $I$ \cite{Nielsen and Chuang 2000}. The Born Rule determines the probability of outcome $i$'s occurence $p_i=\text{Tr}(\rho P_i)$. Quantum state tomography provides a general procedure to estimate an unknown quantum state using measurement outcomes \cite{Nielsen and Chuang 2000}, and more details will be introduced in Subsection \ref{sec3sub1}.

\subsection{Quantum process tomography}\label{qpt}
The framework of quantum process tomography in \cite{Nielsen and Chuang 2000} can be employed to develop identification algorithms for quantum gates. A quantum process $\varepsilon$ maps an input state $\rho_{in}$ to an output state $\rho_{out}$. We let $\rho_{in}$ and $\rho_{out}$ have the same dimension $d$. In Kraus operator-sum representation, we have
\begin{equation}\label{kraus1}
\varepsilon(\rho_{in})=\rho_{out}=\sum_{i}A_i\rho_{in} A_i^\dagger,
\end{equation}
where $A^\dagger$ is the conjugation ($*$) and transposition ($T$) of $A$ and $\{A_i\}$ is a set of mappings (called Kraus operators) from input Hilbert space to output Hilbert space, with $\sum_i A_i^\dagger A_i\leq I$. In this paper we focus on trace-preserving operations, which means satisfying the completeness relation
\begin{equation}\label{trpreserve}
\sum_i A_i^\dagger A_i= I.
\end{equation}
By expanding $\{A_i\}$ in a fixed family of basis matrices $\{E_i\}$ (not necessarily Hermitian), we obtain $A_i=\sum_j c_{ij}E_j$, and $$\varepsilon(\rho_{in})=\sum_{jk}E_j\rho_{in} E_k^\dagger x_{jk},$$ with $x_{jk}=\sum_i c_{ij}c_{ik}^*$. Denote matrices $C=[c_{ij}]$ and $X=[x_{ij}]$, we have $X=C^TC^*$. Hence, $X$ is Hermitian and positive semidefinite. $X$ is called {\itshape process matrix} \cite{brien 2004}. $X$ and $\varepsilon$ are one-to-one correspondent. Hence, by reconstructing $X$ we obtain the full characterization of $\varepsilon$. The completeness constraint (\ref{trpreserve}) now is
\begin{equation}\label{ptrace1}
\sum_{j,k}x_{jk}E_k^{\dagger}E_j=I,
\end{equation}
which is difficult to be further simplified before the structure of $\{E_i\}$ is determined. Using Lemma \ref{ptracepropo} in Appendix \ref{app1}, we can simplify (\ref{ptrace1}) as $\text{Tr}_1X=I_d$ when $\{E_i\}$ is the natural basis $\{|j\rangle\langle k|\}_{1\leq j,k\leq d}$ (see, e.g., \cite{my 2016}), where $\text{Tr}_1(\cdot)$ denotes the partial trace on $\mathbb H_1$ of $\mathbb H_1\otimes\mathbb H_2$, $\otimes$ is the tensor product and $i=(j-1)d+k$. For calculations of the partial trace, see, e.g., \cite{my 2016}.

Let $\{\rho_m\}$ be a complete basis set of the space $\mathbb C_{d\times d}$ consisting of all $d\times d$ complex matrices. If we let $\{\rho_m\}$ be linearly independent, then each output state can be expanded uniquely as
\begin{equation}\label{add1}
\varepsilon(\rho_m)=\sum_n \lambda_{mn}\rho_n.
\end{equation}
Considering the effect of the basis set, we can write
\begin{equation}\label{betadef}
E_j \rho_m E_k^\dagger=\sum_nB_{mn,jk}\rho_n,
\end{equation}
where $B_{mn,jk}$ are complex coefficients. Hence, $$\sum_n\sum_{jk}B_{mn,jk}\rho_n x_{jk}=\sum_n \lambda_{mn}\rho_n.$$ From the linear independence of $\{\rho_m\}$, we have
\begin{equation}\label{ex1}
\sum_{jk}B_{mn,jk} x_{jk}=\lambda_{mn}.
\end{equation}

Let matrix $\Lambda=[\lambda_{mn}]$ and arrange the elements $B_{mn,jk}$ into a matrix $B$:
\begin{equation}\label{matrixB}
B=
\left(\begin{array}{*{6}{c}}
B_{11,11} & B_{11,21} & \cdots & B_{11,12}  & \cdots & B_{11,d^2d^2} \\
B_{21,11} & B_{21,21} & \cdots & B_{21,12}  & \cdots & B_{21,d^2d^2} \\
\multicolumn{6}{c}{\dotfill} \\
B_{12,11} & B_{12,21} & \cdots & B_{12,12}  & \cdots & B_{12,d^2d^2} \\
B_{22,11} & B_{22,21} & \cdots & B_{22,12}  & \cdots & B_{22,d^2d^2} \\
\multicolumn{6}{c}{\dotfill} \\
B_{d^2d^2,11} & B_{d^2d^2,21} & \cdots & B_{d^2d^2,12}  & \cdots & B_{d^2d^2,d^2d^2} \\
\end{array}\right).
\end{equation}

Define the vectorization function as $$\text{vec}(A_{m\times n})= [A_{11},A_{21},...,A_{m1},A_{12},...,A_{m2},...,A_{mn}]^T.$$ We then have
\begin{equation}\label{eq3}
B\text{vec}(X)=\text{vec}(\Lambda).
\end{equation}
Here $B$ is determined once bases $\{E_i\}$ and $\{\rho_m\}$ are chosen, and $\Lambda$ is obtained from experimental data. Denote $\hat X$ as the estimator of $X$. In practice $\hat\Lambda$ always contains noise or uncertainty. Hence, direct inversion or pseudo-inversion of $B$ may fail to generate a physical estimation $\hat X$. A central issue of different QPT protocols (e.g., MLE or BME) is thus to design algorithms to find a physical estimation $\hat X$ such that $B\text{vec}(\hat X)$ is close enough to $\text{vec}(\hat\Lambda)$. The general procedure to deduce Kraus operators $\{\hat A_i\}$ from $\hat X$ is straightforward and one can refer to e.g. \cite{my 2016}.

\subsection{Problem formulation}

We use $U$ to denote a quantum gate. Generally we may have access to the input state $\rho_{in}$ and the output state $\rho_{out}$. The problem is thus to design a series of input states and POVM measurement on the output states and to reconstruct a physical estimation $\hat U$ of the quantum gate $U$. The input state and output state have the following relationship with $U$:
\begin{equation}\label{evo1}
\rho_{out}=U\rho_{in}U^{\dagger},
\end{equation}
where $UU^\dagger=I$. Suppose that the input states and POVM measurements are already determined, one can then either directly reconstruct the gate from POVM measurement results (e.g., MLE method in \cite{jezek 2003}) or first perform quantum state tomography to obtain $\hat\rho_{out}$ and then deduce the quantum gate. In this paper we reconstruct the output states before we identify the quantum gate. In particular, we provide a set of known input states (probe states) $\rho_{in}^{(i)}$ and obtain a set of unknown output states $\rho_{out}^{(i)}$. Using quantum state tomgraphy, we obtain the estimated output states $\hat\rho_{out}^{(i)}$ to reconstruct the quantum gate $\hat U$. A central problem in the gate identification method is to find a unitary $\hat U$ minimizing $\sum_{i}||\hat U\rho_{in}^{(i)}\hat U^{\dagger}-\hat\rho_{out}^{(i)}||^2$, where we use Hilbert-Schimidt norm $||\cdot||$ in this paper.

\section{Quantum gate identification algorithm}\label{Sec3}

\subsection{General framework of the identification algorithm}
We first illustrate the general framework of our gate identification algorithm. As in the QPT procedure of Subsection \ref{qpt}, we choose the input states as a complete basis set of $\mathbb C_{d\times d}$. More specifically, we choose $\rho_{in}$ as natural basis $\{|j\rangle\langle k|\}_{1\leq j,k\leq d}$. The non-Hermitian bases for $j\neq k$ can be obtained as in \cite{Nielsen and Chuang 2000} from
\begin{equation}\label{eq10}
|j\rangle\langle k|=|+\rangle\langle +|+\text{i}|-\rangle\langle -|-\frac{1+\text{i}}{2}|j\rangle\langle j|-\frac{1+\text{i}}{2}|k\rangle\langle k|,
\end{equation}
where $|+\rangle=(|j\rangle+|k\rangle)/\sqrt2$ and $|-\rangle=(|j\rangle+i|k\rangle)/\sqrt2$.

We notice that in this setting all the input states (and all the output states) are pure states. Therefore, we may design a fast quantum state tomography protocol with low computational complexity and employ this protocol to reconstruct $\hat\rho_{out}$. Details about this protocol will be explained in Subsection \ref{sec3sub1}. Then in Subsection \ref{subsec1} we analyze the gate identification problem within the QPT framework and illustrate the procedure to recover $\hat U$ from $\hat\rho_{out}$ with computational complexity $O(d^3)$. The main steps of our identification algorithm are summarized in Subsection \ref{subsec4}. Computational complexity and an analytical error upper bound is given in Section \ref{secadd}.

\subsection{Fast pure-state tomography}\label{sec3sub1}
Quantum state tomography is a procedure that is used to reconstruct an unknown state from POVM measurement results. Existing quantum state tomography methods mainly include Maximum likelihood estimation \cite{hradil 1997}, \cite{paris 2004}, Bayesian mean estimation \cite{paris 2004}, \cite{kohout 2010}, linear regression estimation (LRE) \cite{qibo}, \cite{hou 2016}, \cite{qibo 2017}, etc. For general state tomography with no prior knowledge, the most efficient one among these methods is LRE, which has computational complexity $O(d^4)$ for reconstructing a $d$-dimensional state. Note from (\ref{eq10}) that all the input states are pure. Hence, all the output states from quantum gates are also pure. Using this information we can establish a fast pure-state tomography protocol with computational complexity $O(d^2)$.

Denote an output state to be reconstructed as $\rho$. Under natural basis, $\rho_{ij}=\langle i|\rho|j\rangle$ for $1\leq i,j\leq d$. Since $\rho$ is pure, we can write $\rho=|\psi\rangle\langle\psi|$. Denote the $i$th row and $j$th column of $\rho$ as $\rho_{i\tau}$ and $\rho_{\tau j}$, respectively. We have $\rho_{\tau j}=|\psi\rangle\langle\psi|j\rangle$. $\rho$ is of rank-1 and thus each nonzero column contains all the information apart from a trivial global phase. Therefore, we only need to reconstruct one column of $\rho$. The chosen column should have the largest $|\langle\psi|j\rangle|$ so that the error is suppressed.

First we take the measurement basis as $|j\rangle\langle j|$ ($1\leq j\leq d$) and estimate all $\rho_{jj}=|\langle\psi|j\rangle|^2$, among which the largest one is denoted as $\hat \rho_{ss}$. Note that this index $s$ can not be determined at the beginning of the experiment. Now with $s$ fixed, for each $j\neq s$ we take the measurement basis as $$P_j=\frac{(|s\rangle+|j\rangle)(\langle s|+\langle j|)}{2}$$ and $$Q_j=\frac{(|s\rangle+i|j\rangle)(\langle s|-i\langle j|)}{2}$$ and obtain the corresponding estimators as $\hat\rho_{js}^{+}$ and $\hat\rho_{js}^{-}$. Using (\ref{eq10}) we have $$\hat\rho_{js}=\hat\rho_{js}^{+}+\text{i}\hat\rho_{js}^{-}-\frac{1+\text{i}}{2}\hat\rho_{ss}-\frac{1+\text{i}}{2} \hat\rho_{jj}.$$ Aligning all $\hat\rho_{js}$ we reconstruct $\hat \rho_{\tau s}$, i.e., ${|\hat\psi\rangle\langle\hat\psi|s\rangle}$. Then we take $\hat\rho=\hat \rho_{\tau s}\hat \rho_{\tau s}^\dagger/\text{Tr}(\hat \rho_{\tau s}\hat \rho_{\tau s}^\dagger)$ as the final estimator of $\rho$. It is clear that the above procedure has computational complexity $O(d^2)$.

Now we consider the resource number needed by this protocol. We first prove that for fixed $s$ and $j$, there exists one set of POVM including two elements proportional to $P_j$ and $Q_j$, respectively. To see this, let $$P_j'=\frac{(|s\rangle-|j\rangle)(\langle s|-\langle j|)}{2},$$ $$Q_j'=\frac{(|s\rangle-\text{i}|j\rangle)(\langle s|+\text{i}\langle j|)}{2}$$ and $R_j=\sum_{k\neq j,s}|k\rangle\langle k|$, then $$\frac{1}{2}(P_j+P_j'+Q_j+Q_j')+R_j=I.$$ Therefore, $P_j$ and $Q_j$ can be included in one set of POVM, and the theoretical least number of different sets of POVM needed for this protocol is $1+(d-1)=d$.

In practice it may be more convenient to realize $P_j$ and $Q_j$ in different sets of POVM, which is the scheme we employ in the simulation and experiment part of this paper. Specifically, we notice that all the eigenvalues of $P_j$ are either $1$ or $0$. Therefore, $I-P_j$ is also a positive semidefinite operator. For the same reason, $I-Q_j$ is also positive semidefinite. Hence, we perform two sets of POVM measurements $\{P_j, I-P_j\}$ and $\{Q_j, I-Q_j\}$ for every $j\neq s$. The total number of different POVM sets to reconstruct one output state is $1+2(d-1)=2d-1$.

\subsection{Gate reconstruction from output states}\label{subsec1}

Comparing (\ref{kraus1}) and (\ref{evo1}), it is clear that $U$ is just the Kraus operator. From $A_i=\sum_j c_{ij}E_j$ we know $C$ has only one nonzero row, which indicates $X$ is of rank 1. Let $X=\text{vec}(G)\text{vec}(G)^{\dagger}$ so that the positive semidefinite requirement is naturally satisfied. From now on we assume that $\{E_i\}$ and $\{\rho_i\}$ are both natural bases. Then from Lemma \ref{ptracepropo} we have $$\text{Tr}_1(\text{vec}(G)\text{vec}(G)^{\dagger})=I_d=GG^{\dagger},$$ which means the completeness constraint is equivalent to the requirement that $G$ is unitary. We further have
\begin{equation}
\begin{array}{rl}
&\sum_{i}||\hat U\rho_{in}^{(i)}\hat U^{\dagger}-\hat\rho_{out}^{(i)}||^2\\
=&\sum_{i}||\sum_{jk}E_j \rho_{i}E_k^\dagger\hat x_{jk}-\sum_n\hat\lambda_{in}\rho_n||^2\\
=&\sum_{i}||\sum_{jk}\sum_{n}B_{in,jk}\rho_{n}\hat x_{jk}-\sum_n\hat\lambda_{in}\rho_n||^2\\
=&\sum_i\sum_n|\sum_{jk}B_{in,jk}\hat x_{jk}-\hat\lambda_{in}|^2\\
=&||B\text{vec}(\hat X)-\text{vec}(\hat\Lambda)||^2,\nonumber
\end{array}
\end{equation}
where the relationship in (\ref{betadef}) has been used.

From Lemma \ref{Bsparse} in Appendix \ref{app1} we know under natural basis $B$ is a permutation matrix, which is a special orthogonal matrix such that all elements are $0$ except exactly one $1$ in each column and each row. Using the unitary invariant property of Hilbert-Schimidt norm, we have
\begin{equation}
\begin{split}
&||B\text{vec}(\hat X)-\text{vec}(\hat\Lambda)||\\
=&||\text{vec}(\hat X)-B^T\text{vec}(\hat\Lambda)||\\
=&||\hat X-\text{vec}^{-1}(B^T\text{vec}(\hat\Lambda))||,\nonumber
\end{split}
\end{equation}
where for square $A$ we define $\text{vec}^{-1}[\text{vec}(A)]=A$. Now the central problem in quantum gate identification can be transformed into the following problem:

\begin{prob}\label{problem3}
Given a permutation matrix $B$ and experimental data $\hat\Lambda$, find a unitary matrix $\hat G$ to solve
\begin{equation}
\min_{\hat G} ||\text{vec}(\hat G)\text{vec}(\hat G)^{\dagger}-\text{vec}^{-1}(B^{T}\text{vec}(\hat\Lambda))||.
\end{equation}
\end{prob}

This problem can be solved by two phases. The first phase is to find a $d\times d$ matrix $\hat S$ for
\begin{equation}\label{subp1}
\min_{\hat S}||\text{vec}(\hat S)\text{vec}(\hat S)^{\dagger}-\text{vec}^{-1}(B^{T}\text{vec}(\hat\Lambda))||,
\end{equation}
and the second phase is to find a unitary $\hat G$ for
\begin{equation}\label{subp2}
\min_{\hat G}||\text{vec}(\hat G)-\text{vec}(\hat S)||.
\end{equation}

In the first phase, we denote $$D=\text{vec}^{-1}(B^T\text{vec}(\Lambda)).$$ Since $B$ is a permutation matrix, $\hat D$ will be just a reordering of $\hat\Lambda$'s elements. Hence, there is a one-to-one correspondence between all the elements of $\hat D$ and $\hat\Lambda$. Also the true value $D$ is of rank-one and thus each nonzero row (or column) contains enough (though not all) information about $\text{vec}(G)$. Hence, we can input a subset of input states to the gate $U$ and reconstruct the output states. From these output states we obtain partial elements of $\hat\Lambda$ and determine partial rows of $\hat D$. Then we recover an $\hat S$ from these rows of $\hat D$.

Denote the $((a'-1)d+1)$th row of $\hat D$ ($1\leq a'\leq d$) as $\hat D_{a'\sigma}$. Note that $\hat D_{a'\sigma}$ is different from $\hat D_{a'\tau}$, while the latter denotes the $a'$th row of $\hat D$. From Proposition \ref{prop3} in Appendix \ref{appendprop3} we know under natural basis the $d^3$ elements in $\hat D_{a'\sigma}$ for all the $a'=1,2,...,d$ are just rearranging of the $d^3$ elements in $\hat\Lambda_{ab}$ where $1\leq a\leq d$ and $1\leq b\leq d^2$. Therefore we only need to reconstruct $\varepsilon(|1\rangle\langle k|)$ where $k=1,2,...,d$. From Subsection \ref{sec3sub1} we need to input $3d-2$ classes of probe states to the gate. Since we have explicitly expressed this mapping in Appendix \ref{appendprop3}, it is not necessary to store $B$ and the complexity of calculating $\hat D_{a'\sigma}$ from $\hat\Lambda_{ab}$ is thus only determined by the number of the elements in $\hat\Lambda_{ab}$. The computational complexity is $O(d^3)$.

To solve the problem in (\ref{subp1}), first we calculate all $\hat D_{a'\sigma}$ ($1\leq a'\leq d$) and find the row with the largest row vector norm is $\hat D_{j\sigma}$. For true value we have
\begin{equation}\label{eqad5}
D_{j\sigma}=G_{1j}\text{vec}(G)^\dagger,
\end{equation}
we thus take $\text{vec}(\hat S)=G_{1j}^{-1}\hat D_{j\sigma}^\dagger$, which is
\begin{equation}\label{eqad}
\hat S=G_{1j}^{-1}\text{vec}^{-1}(\hat D_{j\sigma}^\dagger).
\end{equation}
Though we do not know the value of $G_{1j}$, we will later prove that $G_{1j}$ in (\ref{eqad}) can be substituted by any nonzero number. Without loss of generality, we let $G_{1j}=1$ and use
\begin{equation}\label{eqad3}
\hat S=\text{vec}^{-1}(\hat D_{j\sigma}^\dagger)
\end{equation}
instead of (\ref{eqad}) in practical applications.

Now we consider the second phase. Since $$||\text{vec}(\hat G)-\text{vec}(\hat S)||=||\hat G-\hat S||,$$ the problem in (\ref{subp2}) is in essence searching for the best unitary approximant to a given matrix, which is classically solved via matrix polar decomposition (see Theorem IX.7.2 in \cite{bhatia}). If we make singular value decomposition to obtain $\hat S=\hat U_1\hat\Sigma\hat U_2$ with real diagonal $\hat\Sigma$ and unitary $\hat U_1$ and $\hat U_2$ (computational complexity $O(d^3)$ \cite{schur2}), then the optimal solution is
\begin{equation}\label{eqad2}
\hat G=\hat U_1\hat U_2=\hat S|\hat S|^{-1}.
\end{equation}
Using Lemma \ref{unitary1} in Appendix \ref{app1} the final gate estimation is $\hat U=e^{\text{i}\theta}\hat G^T$ where the real number $\theta$ is often called the global phase. This degree of freedom stems in (\ref{evo1}), i.e., $U$ and $e^{\text{i}\theta}U$ will result in exactly the same physical evolution. Therefore, it is natural to resort to other prior knowledge to eliminate this unknown global phase.

Now we consider the effect of $G_{1j}$ when we use (\ref{eqad3}) instead of (\ref{eqad}). If we multiply $\hat S$ with a nonzero real number, then the optimal $\hat G$ from (\ref{eqad2}) does not change. If we multiply $\hat S$ with $e^{\text{i}\theta}$ then this degree of freedom in phase can be incorporated into $\hat U$. This proves that we can substitute $G_{1j}$ in (\ref{eqad}) by any nonzero number, which validates the feasibility of employing (\ref{eqad3}).

\subsection{The general procedure}\label{subsec4}

\begin{figure}
\begin{center}
\includegraphics[width=0.35 \textwidth]{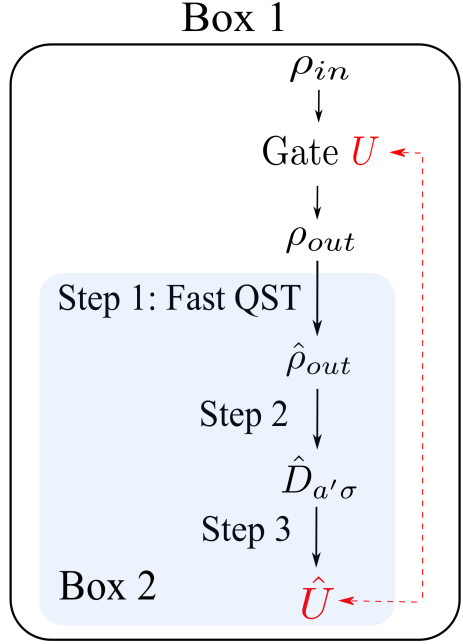}    
\caption{General Procedures for the Gate Identification Algorithm}  
\label{procedures}                                 
\end{center}                                 
\end{figure}

The general procedure of our identification algorithm is summarized in Fig. \ref{procedures}. The three steps in the shaded Box 2 are online computational procedures. When considering the computational complexity, we do not consider experimental procedures outside Box 2.

\textbf{Step 1}. Employ (\ref{eq10}) and the fast quantum state tomography protocol in Subsection \ref{sec3sub1} to reconstruct output states $\hat\varepsilon(|1\rangle\langle k|)$.

\textbf{Step 2}. Based on (\ref{add1}) and Appendix \ref{appendprop3}, obtain $\hat D_{a'\sigma}$ ($1\leq a'\leq d$).

\textbf{Step 3}. For $\hat D_{a'\sigma}$, find the row with the biggest row vector norm as $\hat D_{j\sigma}$. Then $\hat S=\text{vec}^{-1}(\hat D_{j\sigma}^\dagger)$. Let $\hat G=\hat U_1\hat U_2=\hat S|\hat S|^{-1}$, and the estimated gate is $\hat U=\hat G^T$. Use prior knowledge to multiply $\hat U$ with $e^{\text{i}\theta}$ to calibrate the global phase.

\section{Computational complexity and error analysis}\label{secadd}
\subsection{Computational complexity}

For Step 1, there are $3d-2$ classes of output states to be reconstructed. Therefore, the computational complexity is $O(d)\times O(d^2)=O(d^3)$. For Step 2, the procedure to rearrange $\hat\Lambda_{ab}$ into $\hat D_{a'\sigma}$ has computational complexity $O(d^3)$. For Step 3, the calculation of the norms and $\hat S$ is of order $O(d^3)$. The singular value decomposition is also of $O(d^3)$. Hence, the total computational complexity of our algorithm is $O(d^3)$. The complexity is much lower than the $O(d^6)$ in \cite{my 2016}. The reason is that our gate identification method uses only pure states and we develop a fast state tomography algorithm to reconstruct these pure states.

\subsection{Error analysis}

The error in the algorithm has three main resources in this paper: the first one is in POVM measurement, where frequency in simulation or experiment is seldom exactly equal to real probability; the second one comes from state tomography, where the reconstruction of the output states might not be exact; the third one is that the algorithm includes approximation in deduction and might also produce error. Denote $E(\cdot)$ the expectation on all possible measurement outcomes. We here give an analytical error upper bound.

\begin{thm}\label{mainthe}
If $\{E_i\}$ and $\{\rho_m\}$ are chosen as natural basis of $\mathbb{C}_{d\times d}$, then the identification error $E||\hat U-U||$ scales as $O(\frac{d^{2.5}}{\sqrt{N}})$ in the above method, where $N$ is the total number of copies of probe states.
\end{thm}

\begin{pf}

\textit{a) Error in quantum state tomography}

In this paper, we assume to use the same number (denoted as $N_0$) of copies for each input state. Denote the number of different POVM sets as $M$. Then we have $N=(3d-2)N_0$. From the analysis in Subsection \ref{sec3sub1} we know $M\sim O(d)$. Denote $e_{jj}=\hat\rho_{jj}-\rho_{jj}$. According to the central limit theorem $e_{jj}$ converges in distribution to a normal distribution with mean $0$ and variance $\frac{\rho_{jj}-\rho_{jj}^2}{N_0/M}$. For the same reason, $e_{js}^{+}$ and $e_{js}^{-}$ converge to zero-mean normal distributions with variances $\frac{\rho_{js}^{+}-(\rho_{js}^{+})^2}{N_0/M}$ and $\frac{\rho_{js}^{-}-(\rho_{js}^{-})^2}{N_0/M}$, respectively.

In the asymptotical sense the identification error is small enough and we can ensure that $\hat\rho_{ss}$ is close enough to the largest diagonal element of $\rho$ (though it might not be $\rho_{ss}$). Therefore, we may have $1\geq\rho_{ss}>0$ and $\rho_{ss}>\frac{\rho_{jj}}{2}$ for every $j\neq s$. Hence, we know $$\frac{E(e_{ss}^2)}{\rho_{ss}}=\frac{\rho_{ss}-\rho_{ss}^2}{\rho_{ss}N_0/M}\leq\frac{M}{N_0}.$$ For $j\neq s$, $$\frac{E(e_{jj}^2)}{\rho_{ss}}=\frac{\rho_{jj}-\rho_{jj}^2}{\rho_{ss}N_0/M}< \frac{2(1-\rho_{jj})}{N_0/M}\leq \frac{2M}{N_0}.$$

Decompose $\rho=L^\dagger L$. From Cauchy inequality, we have $$|\rho_{js}|^2=|\langle Lj|Ls\rangle|^2\leq\langle Lj|Lj\rangle\langle Ls|Ls\rangle=\rho_{jj}\rho_{ss}.$$ Thus the following relationship holds:
\begin{equation}
\begin{array}{rl}
\rho_{js}^{+}&=\frac{1}{2}\text{Tr}[\rho(|s\rangle+|j\rangle)(\langle s|+\langle j|)]\\
&=\frac{1}{2}(\rho_{ss}+\rho_{jj}+2\Re\rho_{js})\\
&\leq\frac{1}{2}(\rho_{ss}+\rho_{jj}+ 2\sqrt{\rho_{jj}\rho_{ss}})\\
&\leq\frac{3+2\sqrt{2}}{2}\rho_{ss}, \\
\end{array}
\end{equation}
where $\Re(x)$ returns the real part of $x$.
Therefore,
$$\frac{E[(e_{js}^{+})^2]}{\rho_{ss}}= \frac{\rho_{js}^{+}-(\rho_{js}^{+})2}{\rho_{ss}N_0/M}\leq\frac{3+2\sqrt{2}}{2}\frac{(1-\rho_{js}^{+})}{N_0/M}\leq \frac{3M}{N_0}.$$
Similarly $$\frac{E[(e_{js}^{-})^2]}{\rho_{ss}}\leq \frac{3M}{N_0}.$$

Using (\ref{eq10}), we have
\begin{equation}
\begin{array}{rl}
&E(|\hat\rho_{js}-\rho_{js}|^2)\\
=&E[(e_{js}^{+}-\frac{1}{2}e_{jj}-\frac{1}{2}e_{ss})^2]+E[(e_{js}^{-} -\frac{1}{2}e_{jj}-\frac{1}{2}e_{ss})^2]\\
=&E[(e_{js}^{+})^2]+E[(e_{js}^{-})^2]+\frac{1}{2}E(e_{jj}^2)+\frac{1}{2}E(e_{ss}^2)-E(e_{js}^{+}e_{jj})\\
&\ \ \ \ -E(e_{js}^{+}e_{ss})-E(e_{js}^{-}e_{jj})- E(e_{js}^{-}e_{ss})-E(e_{jj}e_{ss})\\
\leq& E[(e_{js}^{+})^2]+E[(e_{js}^{-})^2]+\frac{1}{2}E(e_{jj}^2)+\frac{1}{2}E(e_{ss}^2)\\
&\ \ \ \ +\sqrt{E[(e_{js}^{+})^2]E[(e_{jj})^2]}+\sqrt{E[(e_{js}^{+})^2]E[(e_{ss})^2]}\\
&\ \ \ \  +\sqrt{E[(e_{js}^{-})^2]E[(e_{jj})^2]}+\sqrt{E[(e_{js}^{-})^2]E[(e_{ss})^2]}\\
&\ \ \ \ +\sqrt{E[(e_{jj})^2]E[(e_{ss})^2]}\\
=&E[(e_{js}^{+})^2]+E[(e_{js}^{-})^2]+\frac{1}{2}E(e_{jj}^2)+\frac{1}{2}E(e_{ss}^2)\\
&\ \ \ \ +\{\sqrt{E[(e_{js}^{+})^2]}+\sqrt{E[(e_{js}^{-})^2]}\}\{\sqrt{E[(e_{jj})^2]}\\
&\ \ \ \ +\sqrt{E[(e_{ss})^2]}\}+\sqrt{E[(e_{jj})^2]E[(e_{ss})^2]}\\
\leq& [\frac{3M}{N_0}+\frac{3M}{N_0}+\frac{M}{N_0}+\frac{M}{2N_0}\\
&\ \ \ \ +(\sqrt{\frac{3M}{N_0}}+\sqrt{\frac{3M}{N_0}}) (\sqrt{\frac{2M}{N_0}}+\sqrt{\frac{M}{N_0}})+\frac{\sqrt{2}M}{N_0}]\rho_{ss}\\
\leq&\frac{18M}{N_0}\rho_{ss}.\\  \nonumber
\end{array}
\end{equation}

Then the following relationship holds:
\begin{equation}\label{eqa1}
\begin{array}{rl}
&E(||\hat\rho_{\tau s}-\rho_{\tau s}||^2)/\rho_{ss}\\
&=E(e_{ss}^2)/\rho_{ss}+\sum_{j=1,j\neq s}^d E(|\hat\rho_{js}-\rho_{js}|^2)/\rho_{ss}\\
&\leq (18d-17)\frac{M}{N_0}.\\
\end{array}
\end{equation}

Denote $\Delta_{st}=||\hat\rho_{out}-\rho_{out}||$. In the following we label other errors as the form of $\Delta$ with a subscript. Now we use Lemma \ref{lemma2} in Appendix \ref{app1} to obtain
\begin{equation}\label{eqa2}
\begin{array}{rl}
&\Delta_{st}=||\hat\rho_{out}-\rho_{out}||=||\frac{\hat\rho_{\tau s}\hat\rho_{\tau s}^\dagger} {||\hat\rho_{\tau s}||^2}-\frac{\rho_{\tau s}\rho_{\tau s}^\dagger}{||\rho_{\tau s}||^2}||\\
\leq&(||\frac{\hat\rho_{\tau s}}{||\hat\rho_{\tau s}||}||+||\frac{\rho_{\tau s}}{||\rho_{\tau s}||}||)||\frac{\hat\rho_{\tau s}}{||\hat\rho_{\tau s}||} -\frac{\rho_{\tau s}}{||\rho_{\tau s}||}||\\
=&2|| \frac{\hat\rho_{\tau s}}{||\hat\rho_{\tau s}||}-\frac{\rho_{\tau s}}{||\rho_{\tau s}||}||\\
\leq&2|| \frac{\hat\rho_{\tau s}}{||\hat\rho_{\tau s}||}-\frac{\hat\rho_{\tau s}}{||\rho_{\tau s}||}||+2|| \frac{\hat\rho_{\tau s}}{||\rho_{\tau s}||}-\frac{\rho_{\tau s}}{||\rho_{\tau s}||}||\\
=&2\frac{|||\hat\rho_{\tau s}||-||\rho_{\tau s}|||}{||\rho_{\tau s}||} +2\frac{||\hat\rho_{\tau s}-\rho_{\tau s}||} {||\rho_{\tau s}||}\\
\leq& 4\frac{||\hat\rho_{\tau s}-\rho_{\tau s}||}{||\rho_{\tau s}||}=4\frac{||\hat\rho_{\tau s}-\rho_{\tau s}||}{|\langle\psi|s\rangle|}
=4\frac{||\hat\rho_{\tau s}-\rho_{\tau s}||} {\sqrt{\rho_{ss}}}.\\
\end{array}
\end{equation}

We establish an upper bound (in the asymptotical sense) for the MSE of fast pure-state tomography protocol
\begin{equation}\label{deltast}
\begin{array}{rl}
E\Delta_{st}^2&\leq 16E(||\hat\rho_{\tau s}-\rho_{\tau s}||^2)/\rho_{ss}\\
&\leq16(18d-17)\frac{M}{N_0}\sim O(\frac{d^2}{N_0})\sim O(\frac{d^3}{N}).\\
\end{array}
\end{equation}

For input basis matrices $\{\rho_m\}$, when $\rho_m$ is Hermitian, it corresponds to just one probe state. Hence, $$E||{\hat\varepsilon(\rho_m)}-\varepsilon(\rho_m)||^2\leq E\Delta_{st}^2.$$ When $\rho_m$ is not Hermitian, it is in fact a linear combination of four probe states according to (\ref{eq10}). For $j\neq k$ we have
\begin{equation}
\begin{array}{rl}
&||{\hat\varepsilon(|j\rangle\langle k|)}-\varepsilon(|j\rangle\langle k|)||\\
 =& ||[{\hat\varepsilon(|+\rangle\langle +|)}-\varepsilon(|+\rangle\langle +|)]+\text{i}[{\hat\varepsilon(|-\rangle\langle -|)}-\varepsilon(|-\rangle\langle -|)]\\
&\ \ \ \ -\frac{1+\text{i}}{2}[{\hat\varepsilon(|j\rangle\langle j|)}-\varepsilon(|j\rangle\langle j|)]\\
&\ \ \ \ -\frac{1+\text{i}}{2}[{\hat\varepsilon(|k\rangle\langle k|)}-\varepsilon(|k\rangle\langle k|)]||\\
\leq & (1+|\text{i}|+|\frac{1+\text{i}}{2}|+|\frac{1+\text{i}}{2}|)\Delta_{st}\sim O(\Delta_{st}).\\ \nonumber
\end{array}
\end{equation}

Therefore for each input basis matrix $\rho_m$, $$||{\hat\varepsilon(\rho_m)}-\varepsilon(\rho_m)||\sim O(\Delta_{st}).$$

\textit{b) Error in $G$}

We denote $w=G_{1j}^*$, which is a number dependent only on the real gate. Since the global phase in $G$ can be eliminated via prior knowledge, we assume that $w$ is real and positive. From (\ref{eqad5}) and (\ref{eqad3}) we know $S=G_{1j}^*G=wG$. To estimate $||\hat G-G||$, we use
\begin{equation}\label{eqtemx}
\begin{array}{rl}
||\hat G-G||&\leq||\hat G-\frac{\hat S}{w}||+||\frac{\hat S}{w}-\frac{S}{w}||+||\frac{S}{w}-G||\\
&=||\hat G-\frac{\hat S}{w}||+||\frac{\hat S}{w}-\frac{S}{w}||,\\
\end{array}
\end{equation}
where $||\hat G-\frac{\hat S}{w}||$, $|w|$ and $||\hat S||$ will be separately estimated below.

\textit{c) Error in $S$}

Denote $\Delta_\Lambda=||\hat\Lambda-\Lambda||$, from the analysis in \cite{my 2016} we know
\begin{equation}\label{lambupper}
\begin{array}{rl}
\Delta_\Lambda^2 &=\sum_{m=1}^{d}||\hat{\varepsilon}(\rho_m)-\varepsilon(\rho_m)||^2\sim O(\Delta_{st}^2)\\
\end{array}
\end{equation}
and
\begin{equation}\label{eqtemx1}
||\hat D-D||=\Delta_\Lambda.
\end{equation}

We thus have
\begin{equation}\label{eqtem12}
\begin{array}{rl}
||\hat S-S||&=||\text{vec}(\hat S)^\dagger-\text{vec}(S)^\dagger||=||\hat D_{j\sigma}-D_{j\sigma}||\\
&\leq||\hat D-D||=\Delta_\Lambda\\
\end{array}
\end{equation}

\textit{d) Estimation of $||\hat G-\frac{\hat S}{w}||$}

For the true value we have $$S^\dagger S=|w|^2I, ||S||=|w|\sqrt{d}.$$ Denote $$||\hat S^\dagger \hat S-S^\dagger S||/|w|^2=\Delta_{\Gamma}.$$ Perform spectral decomposition $\hat S^\dagger\hat S/|w|^2=\hat U\hat E\hat U^\dagger$, where $$\hat E=\text{diag}(1+t_1,1+t_2,...1+t_d)$$ and $t_j\in\mathbb R$. Then
\begin{equation}
\begin{split}
\Delta_{\Gamma}^2=&||\hat S^\dagger \hat S-S^\dagger S||^2/|w|^4\\
=&||\hat U\hat E\hat U^\dagger-I||^2=||\hat E-I||^2\\
=&\sum_j t_j^2.  \nonumber
\end{split}
\end{equation}
Thus we know
\begin{equation}\label{eqtem6}
\begin{array}{rl}
&||\hat G-\frac{\hat S}{w}||^2\\
=&\text{Tr}[(\hat G^\dagger- \frac{\hat S^\dagger}{w^*})(\hat G- \frac{\hat S}{w})]\\
=&\text{Tr}(\hat G^\dagger\hat G)-\text{Tr}(\frac{\hat S^\dagger}{w^*}\hat G+\hat G^\dagger\frac{\hat S}{w})+\text{Tr}(\frac{\hat S^\dagger\hat S}{|w|^2})\\
=&d-2\text{Tr}\sqrt{\hat S^\dagger \hat S}/|w|+\text{Tr}(\hat S^\dagger \hat S)/|w|^2\\
=&d-2\sum_j\sqrt{1+t_j}+\sum_j(1+t_j)\\
=&\sum_j(\sqrt{1+t_j}-1)^2=\sum_j\frac{t_j^2}{2+t_j+2\sqrt{1+t_j}}\\
=&\sum_j t_j^2[\frac{1}{4}-\frac{1}{8}t_j+o(t_j)]=\frac{1}{4}\Delta_{\Gamma}^2+o(\Delta_{\Gamma}^2).\\
\end{array}
\end{equation}

For $\Delta_{\Gamma}$, we have
\begin{equation}\label{eqtem3}
\begin{array}{rl}
&\Delta_{\Gamma} =||\hat S^\dagger \hat S-S^\dagger S||/|w|^2\\
\leq&||\hat S^\dagger \hat S-\hat S^\dagger S||/|w|^2+||\hat S^\dagger S- S^\dagger S||/|w|^2\\
\leq&||\hat S^\dagger||/|w|\cdot||\hat S-S||/|w|+||S||/|w|\cdot||\hat S^\dagger- S^\dagger||/|w|\\
=&(||\hat S||/|w|+\sqrt{d})||\hat S-S||/|w|\\
\leq&(||\hat S-S||/|w|+||S||/|w|+\sqrt{d})||\hat S-S||/|w|\\
=&(||\hat S-S||/|w|+2\sqrt{d})||\hat S-S||/|w|\\
\leq&(\Delta_\Lambda/|w|+2\sqrt{d})\Delta_\Lambda/|w|,\\
\end{array}
\end{equation}
where the third line comes from the compatibility of Hilbert-Schimidt norm: $||AB||\leq||A||\cdot||B||$.

Combining (\ref{eqtem3}) and (\ref{eqtem6}), we obtain
\begin{equation}\label{eqtem11}
\begin{array}{rl}
||\hat G-\frac{\hat S}{w}||&=\sqrt{\frac{1}{4}\Delta_{\Gamma}^2+o(\Delta_{\Gamma}^2)}\\
&\leq \sqrt{(\frac{\Delta_\Lambda^2}{|w|^2}+4d+4\sqrt{d}\frac{\Delta_\Lambda}{|w|}) \frac{\Delta_\Lambda^2}{4|w|^2}+o(\Delta_\Lambda^2)}\\
&=\sqrt{\frac{d\Delta_\Lambda^2}{|w|^2}+o(\Delta_\Lambda^2)}\\
&=\sqrt{d}\frac{\Delta_\Lambda}{|w|}+o(\Delta_\Lambda)\\
\end{array}
\end{equation}

\textit{e) Estimation of $|w|$}

In the asymptotic sense the identification error will be small enough. Hence, $|w|$ will be close enough to $\max_{k}|G_{1k}|$. Since $G$ is unitary, each row is a unit vector and $$\max_{k}|G_{1k}|\geq\frac{1}{\sqrt{d}}.$$ Therefore, we can have
\begin{equation}\label{eq31}
|w|\geq\frac{1}{2\sqrt{d}}.
\end{equation}

\textit{f) Total error}\label{subsectotalerror}
Now we substitute (\ref{eqtem12}) and (\ref{eqtem11}) into (\ref{eqtemx}) and employ (\ref{eq31}) and (\ref{lambupper}) to obtain
\begin{equation}\label{eqtem13}
\begin{array}{rl}
||\hat G-G||&\leq||\hat G-\frac{\hat S}{w}||+||\frac{\hat S}{w}-\frac{S}{w}||\\
&\leq\sqrt{d}\frac{\Delta_\Lambda}{|w|}+\frac{\Delta_\Lambda}{|w|}+o(\Delta_\Lambda)\\
&\sim O(d\Delta_\Lambda)\sim O(d\Delta_{st}).
\end{array}
\end{equation}
Using (\ref{deltast}) and taking expectation, we have $$E||\hat G-G||\sim O(dE\Delta_{st})\sim O(\frac{d^{2.5}}{\sqrt{N}}).$$ Finally from Lemma \ref{unitary1} we have
\begin{equation}
E||\hat U-U||\sim O(\frac{d^{2.5}}{\sqrt{N}}).
\end{equation}
\hfill $\Box$
\end{pf}

\section{Numerical results}\label{Sec4}
\subsection{Error vs resource number}\label{simusec1}

\begin{figure}
\begin{center}
\includegraphics[width=0.7 \textwidth]{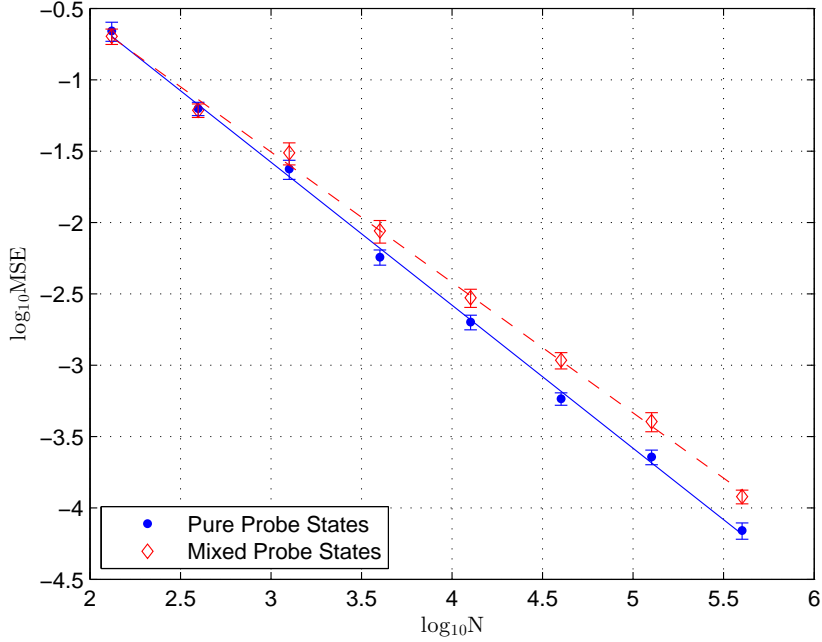}    
\caption{Error MSE versus total resources number $N$}  
\label{test2a}                                 
\end{center}                                 
\end{figure}

We perform numerical simulation to validate Theorem \ref{mainthe} and to showcase the specific performance of the proposed identification algorithm. We consider a single-qubit Hadamard gate,
\begin{equation}\label{ham1}
U=\frac{1}{\sqrt{2}}\left(\begin{array}{*{2}{c}}
1 & 1\\
1 & -1\\
\end{array}\right).
\end{equation}
To compensate the global phase, we assume knowing the prior knowledge that $U_{11}$ is real. Using the identification algorithm in Section \ref{Sec3}, the simulation result is shown in Fig. \ref{test2a}, where the vertical axis is the logarithm of the mean squared error (MSE) (i.e., $\log_{10}E||\hat U-U||^2$) and the horizontal axis is the logarithm of total resource number (i.e., $\log_{10}N$). The blue dots are simulation results with probe pure states and the blue line is the fitting line, with slope $-1.0020\pm0.0150$. Each point is repeated for 50 times. The slope is very close to $-1$, which matches the conclusion in Theorem \ref{mainthe} very well.

We further consider the case when the probe states are not completely pure. We mix each probe pure state with maximally mixed states as $\rho_{in}'=\alpha\rho_{in}+(1-\alpha)\frac{I}{d}$, where $\alpha=\sqrt{0.99}$ and the purity of the mixed states is $\text{Tr}(\rho_{in}'^2)=0.9950$. The simulation results are shown as red diamonds in Fig. \ref{test2a}, and the red fitting line has slope $-0.9141\pm0.0158$. This slope is not far from the theoretical optimal value, which demonstrates that our identification algorithm is applicable even the probe states are not completely pure in the laboratory with current optics technique. From Fig. \ref{test2a}, it is also clear that if the probe states are not completely pure, it is possible to use more copies of mixed states than pure states to achieve a similar level of identification accuracy.

\subsection{Comparison with MLE}

\begin{figure}
\begin{center}
\includegraphics[width=0.65 \textwidth]{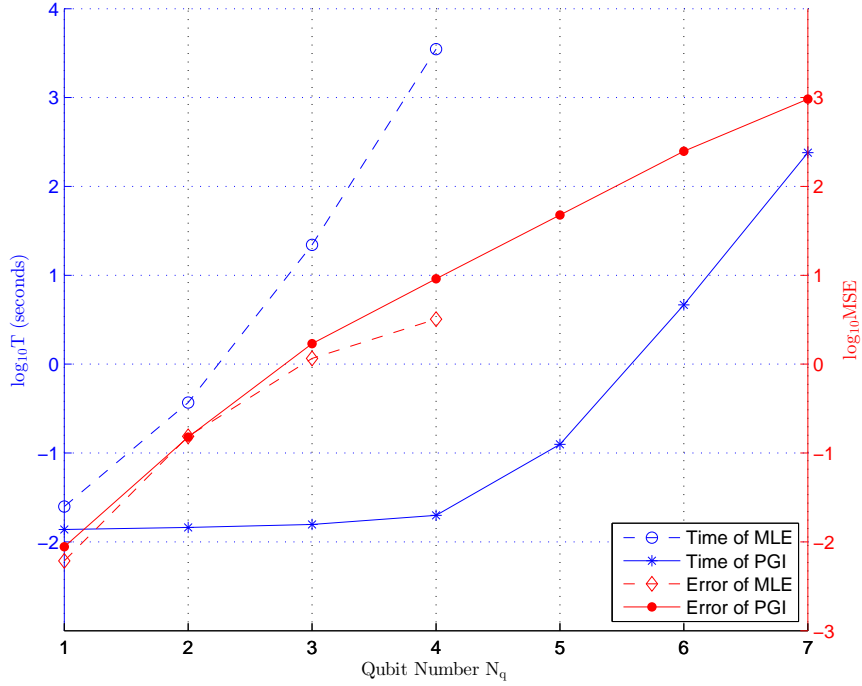}    
\caption{Running time $T$ and MSE versus qubit number $N_q$ for MLE and our Pure-state-based Gate Identification (PGI) method}  
\label{test4}                                 
\end{center}                                 
\end{figure}

We present numerical results to compare the running time and identification error of our algorithm with the Maximum Likelihood Estimation (MLE) method. MLE is the most widely used quantum tomography method. Denote $N_q$ as the number of qubits, and the gate to be identified is in the form of $N_q$ times tensor product of single-qubit Hadamard gate
\begin{equation}
U=\left[\frac{1}{\sqrt{2}}\left(
\begin{array}{cc}
1 & 1\\
1 & -1\\
\end{array}
\right)\right]^{\otimes N_q}.
\end{equation}
We take $N=10^3\times d^2\times(3d-2)$. We perform POVM measurement, reconstruct the output states using the proposed fast QST protocol and identify $U$ using our algorithm. Then we feed the same POVM measurement bases and the corresponding measurement results to MLE algorithm for identification. The simulation result is illustrated in Fig. \ref{test4} and each point is repeated for 10 times. The MLE algorithm in \cite{jezek 2003} is employed. Standard MLE algorithm only reconstructs the unknown process matrix $X$ rather than the unitary gate $U$. Hence, the error we compare in Fig. \ref{test4} is $\log_{10}E||\hat X-X||^2$. The running time ($T$) only includes online computational time. From Fig. \ref{test4} our identification algorithm is much faster than MLE. For example, our algorithm takes less time for a seven-qubit ($d=2^{7}$ dimensional) system than MLE for a four-qubit ($d=2^{4}$ dimensional) system.

\section{Experimental results}\label{SecExp}

\begin{figure}
\begin{center}
\includegraphics[width=0.7 \textwidth]{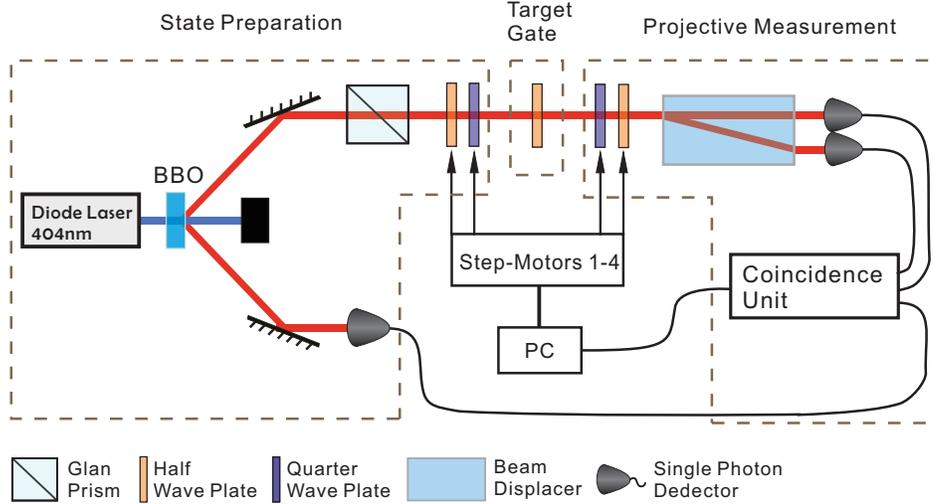}    
\caption{The schematic of experimental setup}  
\label{fig2}                                 
\end{center}                                 
\end{figure}

In the section, we present experimental results on the identification of one-qubit Hadamard gate. The experimental setup is illustrated in Fig. \ref{fig2}. Photon pairs are created using type-I spontaneous parametric down-conversion in a nonlinear crystal. One of the photons is sent immediately to a single photon detector to act as a trigger. The other photon is sent through a Glan prism (extinction ratio more than 2000:1 of horizonal and vertical polarization in the transmission direction) and a half-quarter wave plate combination to prepare it in any desired state of very pure polarization. The Hadamard gate is realized by a half wave plate with its axis placed at $22.5^\circ$ relative to lab horizon. Another quarter-half wave plate combination followed by a beam displacer with high extinction ratio (more than 10000:1) is used to project the photon onto any measurement basis on the Bloch sphere. The rotations of the wave plates in the state preparation part and in the projective measurement part are separately driven by four step-motors, which are connected to a computer with a Labview program to automatically enable the quantum gate identification. Since our method needs to input pure states and assumes output states are also pure, the Glan prism and beam displacer with high extinction ratio are adopted in our experiment to reduce the system error as much as possible, which is measured about 2000:1 for both horizonal and vertical polarization for the whole setup. To alleviate the drift of the collective efficiency of two photon detectors behind the beam displacer, multimode fibers fully covered by black plastic bags instead of singlemode fibers are used to collect the coupled photons. Because of the introducing of multimode fibers we set the coincidence window to $1ns$ to minimize the random coincidence count so that its error is negligible.

\begin{figure}
\begin{center}
\includegraphics[width=1 \textwidth]{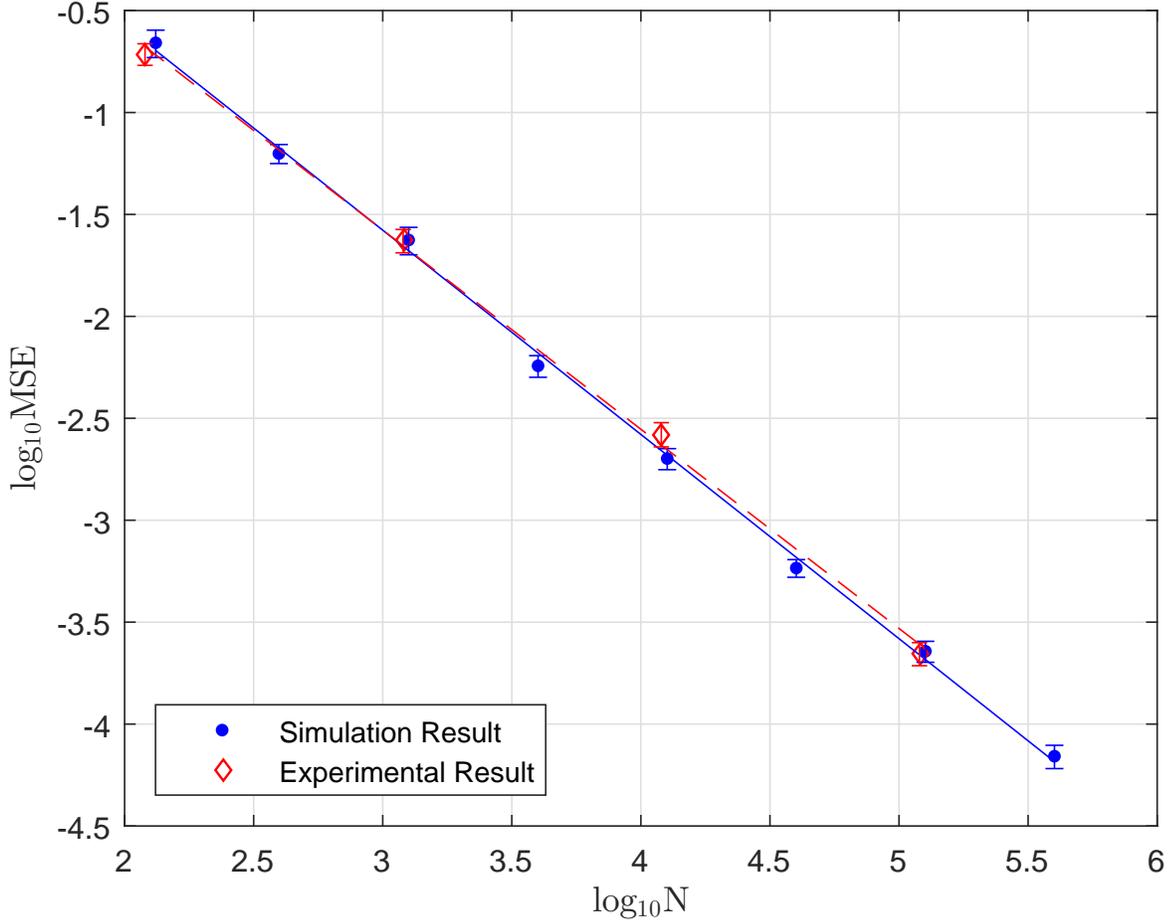}    
\caption{Error (MSE) versus total resource number $N$ for the experimental single-qubit gate}  
\label{test2}                                 
\end{center}                                 
\end{figure}

We generate a quantum gate close to the Hadamard gate, and use our identification algorithm to calibrate it with $N=4\times 10^6$ resources. The identification result is taken as the real value of the gate to be identified later. Then we experimentally identify it with different numbers of resources far less than $4\times10^6$ using our method. The result is shown in red diamonds in Fig. \ref{test2} and every point is averaged over 50 experimental runs. In comparison, we also use blue dots to represent the simulation result of single-qubit Hadamard gate using probe pure states. The fitting line of the experimental result has slope $-0.9772\pm0.0296$, which matches the theoretical result very well.


\section{Conclusion}\label{secfinal}
We have proposed a new method using input pure states to identify an unknown quantum gate. The method is presented within the QPT framework and has computational complexity $O(d^3)$ with the system dimension $d$. In our method, no restriction is placed on the unitary gate to be identified. The algorithm has potential for parallel processing in that during the Step 1 one can deal with available data to reconstruct existing output states while at the same time input new probe states to the gate and make measurement on them. We also established an error upper bound $O(\frac{d^{2.5}}{\sqrt{N}})$ with $N$ the number of copies of all the probe states. The computational complexity and the characterization of error can be useful for designing gate-related experiments or simulation tasks. We performed simulation to compare our algorithm with MLE, which demonstrates the efficiency of our method. We performed a quantum optical experiment on one-qubit Hadamard gate to illustrate the effectiveness of the proposed identification method.

\appendix

\section{Several Lemmas}\label{app1}
\begin{lem}[\cite{my 2016}]\label{ptracepropo}
If $\{E_i\}$ is chosen as $\{|j\rangle\langle k|\}_{1\leq j,k\leq d}$ (called natural basis) and the relationship between $i$ and $j,k$ is $i=(j-1)d+k$, then the completeness constraint is $\textup{Tr}_1X=I_d$.
\end{lem}

\begin{lem}[\cite{my 2016}]\label{unitary1}
Suppose under natural basis assumption we have obtained a unitary estimator $\hat G$. Then there is a Kraus representation where only one operator $\hat A$ is nonzero. $\hat A$ must be unitary and in fact equals to $e^{\textup{i}\phi}\hat G^T$, where $\phi\in\mathbb{R}$.
\end{lem}

\begin{lem}[\cite{my 2016}]\label{Bsparse}
Let $\{E_i\}_{i=1}^{d^2}$ be a set of matrices in $\mathbb C_{d\times d}$. Choose $\{\rho_m\}$ as the natural basis. Then $\{E_i\}=e^{\textup{i}\theta}\{\rho_m\}$ if and only if $B$ is a permutation matrix. Here $\theta \in \mathbb{R}$ is any fixed global phase.
\end{lem}

\begin{lem}[\cite{my 2016}]\label{lemma2}
Let $b$ and $c$ be two complex vectors with the same finite dimension. Then we have
\begin{equation}
||bb^\dagger-cc^\dagger||\leq(||b||+||c||)||b-c||.
\end{equation}
\end{lem}

\section{Proposition \ref{prop3} and its proof}\label{appendprop3}

\begin{prop}\label{prop3}
If $\{E_i\}$ and $\{\rho_m\}$ are natural basis, then the following two sets are equal:  $\{\hat\Lambda_{ab}|1\leq a\leq d,1\leq b\leq d^2\}=\{\hat D_{b'c'}|b'=(a'-1)d+1,1\leq a'\leq d,1\leq c'\leq d^2\}$.
\end{prop}

\begin{pf}
Since $\hat D=\text{vec}^{-1}[B^T\text{vec}(\hat\Lambda)]$, we have
\begin{equation}\label{eqad7}
B^T\text{vec}(\hat\Lambda)=\text{vec}(\hat D).
\end{equation}
From the equivalence of (\ref{ex1}) and (\ref{eq3}), we know (\ref{eqad7}) is equivalent to
\begin{equation}\label{eqb1}
\sum_{ab}B_{ab,jk}\hat\Lambda_{ab}=\hat D_{jk}.
\end{equation}

When calculating the summation in \ref{eqb1}, the indices $a$ and $b$ run from $1$ to $d^2$, and exactly one column ($(jk)-th$ column, which has altogether $d^4$ elements) of $B$ is employed. Since $B$ is a permutation matrix, among these $d^4$ elements only one element is $1$ and all the other elements are $0$. Therefore, each $\hat D_{jk}$ equals exactly to one element of $\hat\Lambda$ and receives no influence from all the other elements of $\hat\Lambda$. The crucial problem is to find this nonzero element in the $(jk)-th$ column of $B$.

Denote by $\delta$ the Dirac Delta function. From (29) in \cite{my 2016},
\begin{equation}\label{eqb2}
\delta_{qg}\delta_{pm}\delta_{th}\delta_{sn} =B_{(s,t)(p,q),{(m,n)(g,h)}}.
\end{equation}

In (\ref{eqb2}), the notation $(x,y)$ is short for the number $(x-1)d+y$ with $1\leq x,y\leq d$. To find the nonzero element in each column of $B$, we substitute $g$, $m$, $h$ and $n$ into $q$, $p$, $t$ and $s$ respectively. The RHS of (\ref{eqb2}) becomes
\begin{equation}\label{eqb3}
B_{(s,t)(p,q),(p,s)(q,t)},
\end{equation}
which are all the nonzero elements in $B$. Comparing (\ref{eqb3}) with (\ref{eqb1}), we have
\begin{equation}\label{eqb4}
\left\{\begin{array}{rl}
a&=(s,t)\\
b&=(p,q)\\
j&=(p,s)\\
k&=(q,t)\\
\end{array}\right.
\end{equation}

In (\ref{eqb4}), the dominant indices are $a$ and $b$, which change from $1$ to $d$ and $1$ to $d^2$, respectively. Then we consider the subordinate indices $s$, $t$, $p$, $q$, $j$ and $k$. From the first equation of (\ref{eqb4}), $a=(s-1)d+t$, we know \begin{equation}\label{eqb5}
s\equiv 1
\end{equation}
and
\begin{equation}\label{eqb6}
t=a.
\end{equation}
From the second equation of (\ref{eqb4}), $b=(p-1)d+q$, we have
\begin{equation}\label{eqb7}
q=1+(b-1)\bmod d
\end{equation}
and
\begin{equation}\label{eqb8}
p=\{b-[1+(b-1)\bmod d]\}/d+1.
\end{equation}

(\ref{eqb6})-(\ref{eqb8}) indicates that $t$, $p$ and $q$ will all change from $1$ to $d$. From the third equation in (\ref{eqb4}), we have $j=(p-1)d+1$ for $p=1,2,...,d$. From the fourth equation in (\ref{eqb4}), $k$ will run from $1$ to $d^2$, which proves the conclusion.

Furthermore, we can write the specific mapping by cancelling $s$, $t$, $p$ and $q$ from (\ref{eqb4}) to (\ref{eqb8}). $\hat\Lambda_{ab}$ is now in the $[b-(b-1)\bmod d]$th row and $\{[(b-1)\bmod d]d+a\}$th column of $\hat D$.
\hfill $\Box$
\end{pf}


\end{document}